\documentclass[twocolumn,showpacs,preprintnumbers,amsmath,amssymb]{revtex4}

\usepackage{graphicx}
\usepackage{dcolumn}
\usepackage{bm}

\begin{document}
\def\a{\alpha}
\def\b{\beta}
\def\g{\gamma}
\def\G{\Gamma}
\def\d{\delta}
\def\D{\Delta}
\def\e{\epsilon}
\def\k{\kappa}
\def\l{\lambda}
\def\s{\sigma}
\def\t{\tau}
\def\om{\omega}
\def\Om{\Omega}
\def\lg{\langle}
\def\rg{\rangle}
\newcommand{\be} {\begin{equation}}
\newcommand{\ee} {\end{equation}}
\newcommand{\Be} {\begin{eqnarray}}
\newcommand{\Ee} {\end{eqnarray}}

\preprint{APS/123}
\title{Rotational Correlation Functions of Single Molecules}

\author{G. Hinze}
 \email{hinze@mail.uni-mainz.de}
 \homepage{http://www.uni-mainz.de/FB/Chemie/Basche/}
\author{G. Diezemann}
\author{Th.{ }Basch\'{e}   }
\affiliation{Institut f\"ur  Physikalische Chemie,
 Johannes Gutenberg-Universit\"at, D-55099 Mainz}
\date{\today}

\begin{abstract}
Single molecule rotational correlation functions are analyzed for
several reorientation geometries. Even for the simplest model of
isotropic rotational diffusion our findings predict
non-exponential correlation functions to be observed by
polarization sensitive single molecule fluorescence microscopy.
This may have a deep impact on interpreting the results of
molecular reorientation measurements in heterogeneous
environments.
\end{abstract}

\pacs{33.15.Vb, 87.64.Ni, 61.43.FS, 67.40.Fd}

\maketitle

Reorientation of small molecules or segments of macromolecules
undergoing conformational changes are elementary processes and
often of crucial importance for the properties of materials. In
most of the experimental techniques probing rotational dynamics,
ensembles of molecules have been monitored
\cite{berne1990,boettcher1992,boehmer2001}. Actually, many
chemical and biological systems are heterogeneous, a fact that
e.g. is reflected in their dynamical properties. In the context of
the non-exponential decay of bulk correlations in supercooled
liquids this has been termed dynamic heterogeneity. Proteins,
polymers, colloids and coarsening systems share this feature of
non-exponential relaxation.
In the past, bulk techniques have been developed that allow a
characterization of heterogeneous rotational dynamics in complex
systems in terms of a distribution of reorientational rates
fluctuating in time\cite{richert2002,sillescu2002}. There are,
however, hints that even without (bulk) heterogeneity rotational
correlations functions (RCFs) show deviations from exponential
decay on a microscopic scale\cite{dreizehn1998}. Such a behavior
has also been found in numerical calculations on spin glasses
\cite{Castillo03,Cugliandolo03}.

In order to get a deeper insight into the nature of the rotational
dynamics and its relation to structural properties of
heterogeneous materials, it is highly desirable to conduct
experiments providing local (as opposed to averaged bulk)
information. It has been shown by numerous examples that a single
molecule is an exquisitely sensitive probe of dynamical processes
in its local environment\cite{basche1997,science1999}. Along these
lines, temporal fluctuations of excited state
lifetimes\cite{vallee2003}, excitation\cite{ambrose1991} and
emission spectra\cite{lu1997} and electronic coupling
strengths\cite{lippitz2004} have been studied by single molecule
spectroscopy and analyzed in terms of fluctuating environments.
Here we focus on the rotational dynamics of single fluorophores
investigated by polarization resolved
microscopy\cite{ha1996,ha1999,harms1999,vdbout1,vdbout2,vdbout3}.
In order to extract quantitative and reliable information from
such measurements, a detailed analysis of local RCFs is of utmost
importance. It is the purpose of the present letter to provide
such an analysis for the simple model of rotational diffusion of
single fluorophores.

Fluorophores preferentially absorb photons whose electric field
vectors are aligned parallel to the transition dipoles of the
molecule \cite{lakowicz1999}. These dipoles have a well defined
orientation with respect to the molecular axes. Similarly,
emission also occurs with light polarized along a fixed axis. The
experimental situation we will analyze in the remainder of this
paper is depicted in Fig. 1 together with the definition of the
axes system and the angles used.
The orthogonal polarization resolved emission signals $I_p$ and
$I_s$ depend on both, absorption and emission efficiency. The
contributions of $I_p$ and $I_s$ to the signals collected in
fluorescence correlation spectroscopy have been analyzed in
detail\cite{pecora75,pecora76}.

{A customary measure of dipole orientation can be obtained by
calculating the reduced linear dichroism $d$ from $I_p$ and $I_s$
via }
\begin{equation}
   d=\frac{I_p-I_s}{I_p+I_s}
\end{equation}
{This frequently used
quantity\cite{harms1999,vdbout1,vdbout2,vdbout3,hochstrasser2003}
fluctuates in the course of time due to molecular rotational
dynamics.}
The normalization ensures that $d$ depends solely on the
orientation of the emission dipole and not on the absorption
efficiency.
{Hence we represent the transition dipoles not by vectors but by
their orientation using angles $\theta, \phi$.
We want to point out that the analogy of $d$ to what is often
called polarization\cite{lakowicz1999} could lead to the wrong
assumption to use the anisotropy values instead. However, in
contrast to fluorescence depolarization methods this would not
make any sense here.}
Subsequently the auto-correlation function of $d(t)$ is
calculated,
\begin{equation}
   C_d(t)=\left< d(t+\tau) d(\tau) \right>
\end{equation}
In the present letter we will show that $C_d(t)$ is a non-trivial
correlation function which nevertheless gives valuable information
about the time scale and the geometric aspects of molecular
reorientations.
{It should be noted that in principle the fraction $I_p/I_s$ leads
to the azimuth $\phi$, however, an extremely well S/N ratio would
be needed for reliable values.}
\begin{figure}
\includegraphics[width=7.5cm,height=3cm]{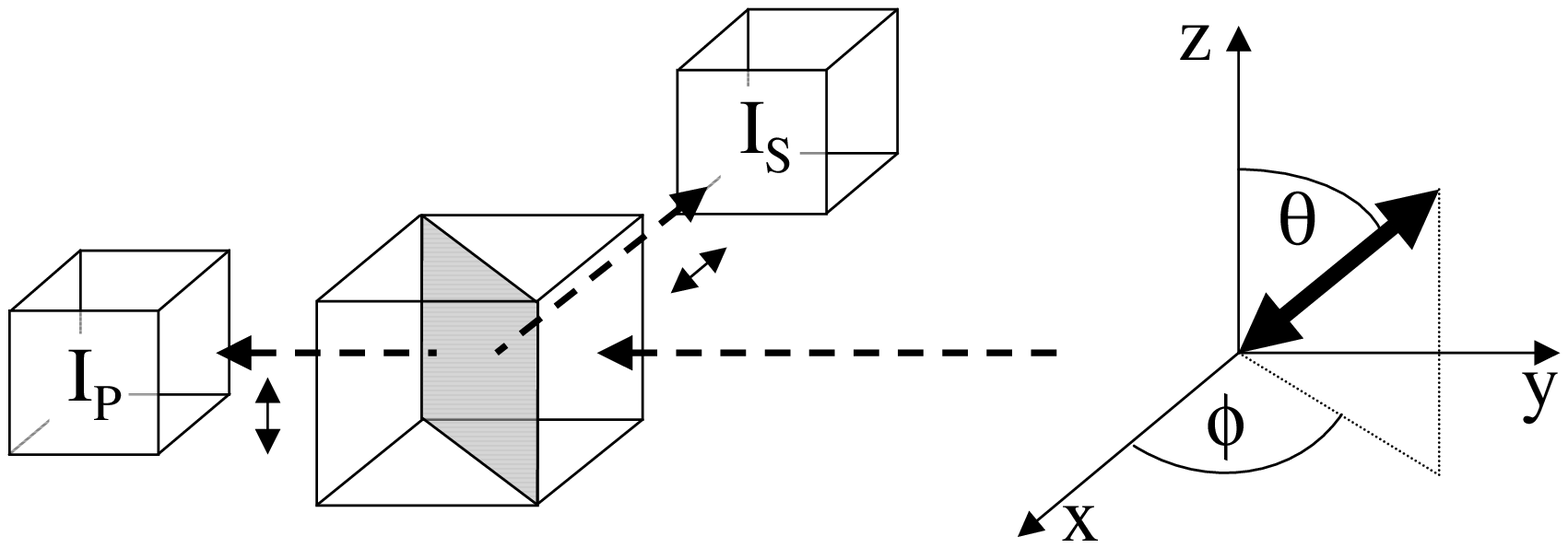}  
\caption{Schematic experimental setup: The polarization of
  light emitted by a single fluorophore depends on the orientation of the transition
  dipoles with respect to the detection system. The two
  orthogonally polarized components of the signal are separated by a polarizing
  cube beam splitter.
  }
  \label{111}
\end{figure}

Well known bulk techniques allowing to monitor molecular
rotational dynamics are provided by dynamic light scattering
\cite{berne1990}, dielectric spectroscopy \cite{boettcher1992} or
nuclear magnetic resonance (NMR) \cite{boehmer2001}. Due to the
different ranks $l$ of the respective interaction tensors distinct
RCFs
\be\label{Cl.t} C_l(t)=\lg
P_l(\cos{(\theta(t))}P_l(\cos{(\theta(0))}\rg \ee
of the relevant Legendre polynomial are obtained. Here, $\theta$
is defined in the same way as in Fig.1. For instance, the
interaction of a permanent electric dipole moment with an applied
electric field transforms in the same way as the first order
Legendre polynomial $P_1(\cos{\theta})=\cos{\theta}$ and therefore
$C_1(t)$ is observed in dielectric spectroscopy. NMR and light
scattering are $l=2$ techniques and yield $C_2(t)$. Below it will
be shown that $C_d(t)$, apart from being a local correlation
function, cannot be related to one of the $C_l(t)$ in a simple
manner.
%
%
%

%
The theoretical description of reorientational dynamics in
molecular liquids is a complicated many particle problem. For a
qualitative understanding of the behavior of rotational
correlation functions, however, a treatment in terms of rotational
Brownian motion is usually sufficient. In the present letter we
restrict ourselves to the simple model of anisotropic rotational
diffusion, thus neglecting inertial effects completely. This
approximation is well suited for the systems we have in mind,
namely biological systems or supercooled liquids and glasses,
where the time scale of the relevant relaxational modes is slow
compared to typical vibrational frequencies.

In all following considerations we assume the transition dipole
matrix elements to be stationary quantities. In this case we have
$C_d(t)=\lg d(t)d(0)\rg$, with
\begin{equation}\label{def_d}
   d(\theta,\phi)={\cos^2{\theta}-\sin^2{\theta}\cos^2{\phi}\over
   \cos^2{\theta}+\sin^2{\theta}\cos^2{\phi}} .
\end{equation}
Note that $d(\theta,\phi)$ cannot be expressed in terms of the
spherical harmonics $Y_{2m}(\theta,\phi)$ and this is why $C_d(t)$
cannot be related to the corresponding RCF $C_2(t)$. This is to be
contrasted to the optical anisotropy, the correlation function of
which is directly proportional to $C_2(t)$.

In order to keep the treatment general, in the following we treat
auto-correlation functions of the form
\be \label{Cx.t} C_X(t)=\lg X(\Om_{PL}(t))X(\Om_{PL}(0))\rg. \ee
Here, $X$ denotes some arbitrary function of the molecular
orientation $\Om_{PL}$ of the principal axis system (P) of the
relevant interaction tensor in the laboratory fixed frame (L). In
the present context we are primarily interested in the case of
fluorescence, $X\!=\!d$, in which case the z-axis of the P-system
coincides with the direction of the transition dipole. However,
other choices like e.g. $X\!=\!e^{i{\bf q}{\bf r}}$ relevant for
incoherent neutron scattering can be analyzed in an identical way.
If the molecular orientation $\Om(t)$, given in terms of Eulerian
angles\cite{rose57}, is modelled as a stochastic process, $C_X(t)$
can be written in the form\cite{vkamp81}:
\be\label{Cx.stoch} C_X(t)={1\over
8\pi^2}\int\!d\Om\int\!d\Om_0X(\Om)X(\Om_0)P(\Om,t|\Om_0) \ee
where $P(\Om,t|\Om_0)$ denotes the conditional probability to find
$\Om$ at time $t$ given $\Om_0$ at time $t\!=\!0$. Throughout this
letter we restrict ourselves to the model of anisotropic
rotational diffusion of symmetric top molecules, in which case one
has
\begin{equation}\label{P.Dlmn}
\begin{array}{c}
  P(\Om,t|\Om_0)=\sum_{l,m,n}\left(\frac{2l+1}{8\pi^2}\right)
             D_{mn}^{(l)*}(\Om_0)D_{mn}^{(l)}(\Om)  \\ \\
   e^{-\left[l(l+1)D_x+m^2(D_z-D_x)\right]t} \\
\end{array}
\end{equation}
Here, the $D_{nm}^{(l)}(\Om)$ are Wigner rotation matrix
elements\cite{rose57}. The rotational diffusion coefficients $D_y$
and $D_x$ are equal, but different from $D_z$. The limit of
isotropic rotational diffusion is recovered for $D_x\!\!=\!\!D_z$.
It is important to point out that $\Om$ in eq.(\ref{P.Dlmn})
denotes the orientation of the diffusion tensor (D) in the
L-system, $\Om\!\equiv\!\Om_{DL}$.

In order to proceed in the calculation of $C_X(t)$, we expand the
quantity $X(\Om_{PL})$ in the expression for the correlation
function, eq.(\ref{Cx.t}), in terms of $D_{nm}^{(l)}$,
$X(\Om_{PL}(t))=\sum_{l,mn}X_{l;mn}D_{nm}^{(l)}(\Om_{PL}(t))$ with
$X_{l;mn}={2l+1\over 8\pi^2}\int\!d\Om D_{nm}^{(l)*}(\Om)X(\Om)$.
Next, $D_{nm}^{(l)}(\Om_{PL}(t))$ is expressed in terms of the
relevant $\Om_{DL}(t)$ via $\Om_{PL}(t)\!=\!\Om_{PD}+\Om_{DL}(t)$
using $D_{nm}^{(l)}(\Om_{PL}(t))\!= \!\sum_\mu
D_{n\mu}^{(l)}(\Om_{PD})D_{\mu m}^{(l)}(\Om_{DL}(t))$. Here,
$\Om_{PD}\!=\!(\a,\theta_{PD},\pi-\phi_{PD})$ is a shape and
symmetry dependent molecular quantity. For most relevant cases
$X(\Om_{PD})$ has axial symmetry, which allows to replace the
$D_{0m}^{(l)}$ by spherical harmonics $Y_{lm}$. Performing the
calculation using eqns.(\ref{Cx.stoch}) and (\ref{P.Dlmn}) yields:
\be\label{Cx.Dlmn} C_X(t)=\sum_l A_l C_l(t) .\ee
With $A_l={1\over 4\pi}
\sum_m\left|\int_0^{2\pi}\!d\phi\int_0^\pi\!d\theta\sin{\theta}
X(\theta,\phi)Y_{lm}(\theta,\phi)\right|^2$ and
\be\label{Fl.def}
C_l(t)=\sum_m\left|D_{0m}^{(l)}(\Om_{PD})\right|^2
     e^{-\left[l(l+1)D_x+m^2(D_z-D_x)\right]t} ,
\ee
the RCFs eq.(\ref{Cx.Dlmn}) are calculated numerically.
Typically, it is sufficient to use $l$ values up to $l\!=\!20$.
Thus, $C_d(t)$ is given by a weighted sum of a large number of
RCFs. This fact has not been noticed in the literature before to
the best of our knowledge. As already mentioned above, for the
optical anisotropy one has $\lg r(t)r(0)\rg\!=\!C_2(t)$ which
decays as a superposition of at most three exponentials.
%
%
%

%
For the case of isotropic rotational diffusion, i.e.
$D=D_x=D_y=D_z$, eq.(\ref{Cx.Dlmn}) simplifies to
\begin{equation} \label{isotrop}
   C_X(t)=\sum_l A_l e^{-l(l+1)Dt}
\end{equation}
For symmetry reasons, the odd components vanish.
{A similar expression has been reported recently
\cite{hochstrasser2003}.}
In table~\ref{tab1} the pre-factors are listed up to $l\!=\!20$.
\begin{table}[h]
\begin{tabular}{|c|c||c|c|}
  \hline
  $l$ & $A_l$ & $l$ & $A_l$ \\
  \hline
  2 & 0.83501300 & 12 & 0.00417137 \\
  4 & 0.10020500 & 14 & 0.00266071 \\
  6 & 0.03101390 & 16 & 0.00180131 \\
  8 & 0.01352020 & 18 & 0.00127513 \\
 10 & 0.00708445 & 20 & 0.00093625 \\
  \hline
\end{tabular}
\caption{Pre-factors $A_l$ from eq.(\ref{isotrop}) for isotropic
rotational diffusion} \label{tab1}
\end{table}
The resulting correlation function significantly deviates from an
exponential decay as shown in Fig.2. From fitting the computed
data with a stretched exponential function
$f(t)=e^{-\left(t/\tau_c\right)^{\beta}}$ we obtain
$\tau_c=0.87/6D$ and $\beta=0.871$. Therefore, even for this
simple model we find intrinsic non-exponential relaxation. The
optical anisotropy decays exponentially, $\lg
r(t)r(0)\rg\!=\!e^{-6Dt}$.
\begin{figure}
  \includegraphics[width=6cm,height=5cm]{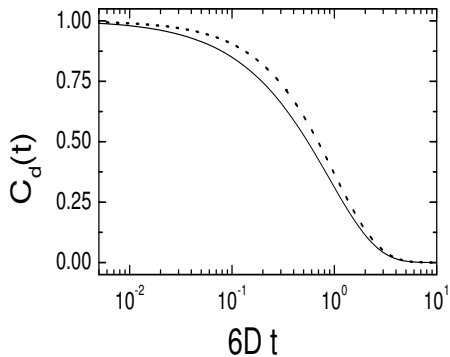}\\
  \caption{Rotational correlation function $C_d(t)$ for isotropic
     diffusion (solid line). For comparison the exponential
     correlation function $l=2$ (eq.\ref{isotrop}:
     $A_2=1, A_{l\neq 2}=0$) is plotted as dotted line}
\label{121}
\end{figure}
%
%
%

\begin{figure}
  \includegraphics[width=7cm,height=9cm]{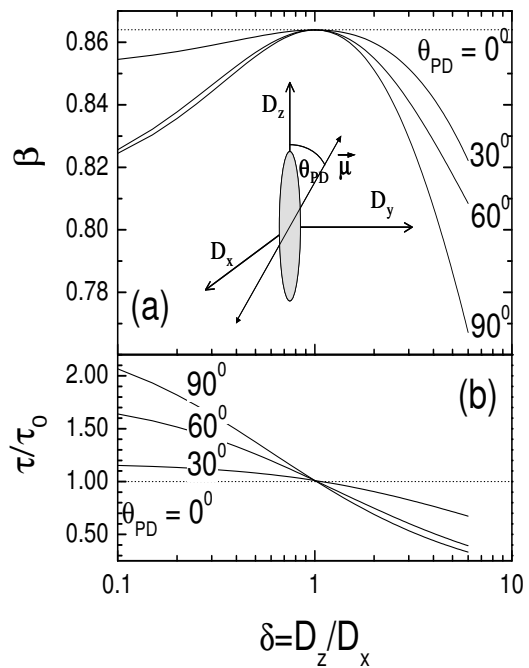}\\
  \caption{(a) Stretching exponent $\beta$ depending on the anisotropy
  $\delta=D_z/D_x$ ($D_x=D_y$). $\theta_{PD}$ denotes the angle between
  the z-axis of the diffusion tensor and the transition dipole $\vec{\mu}$.
  (b) Correlation times normalized by $\tau_0=\tau(D_z\!\parallel\!\vec{\mu})$. $\tau$
  and $\beta$ were obtained by fitting $C(t)$ with
  $exp(-(t/\tau)^{\beta})$
  }
\label{figaniso}
\end{figure}

%
The model of stochastic isotropic reorientations presents a
drastic approximation as most molecules do not exhibit a spherical
shape. Thus, in many cases of practical interest one is confronted
with asymmetric top molecules. As noted above, however, in the
present letter we restrict ourselves to the case of symmetric tops
for simplicity. This means that we consider an ellipsoidal
diffusion tensor with three components $D_x=D_y$ and $D_z$, see
Fig. \ref{figaniso}a. For standard $l=1,2$ methods the RCFs
resulting for this model consist of two and three exponentials,
respectively, with time constants depending on the degree of
anisotropy $\d=D_z/D_x$. It is evident from eq.(\ref{Fl.def}) that
for $D_x\neq D_z$ the degree of non-exponentiality is more
pronounced due to the fact that for each $l$ now one has $l+1$
exponentially decaying functions with different weights. However,
the orientation of the transition dipoles with respect to $D_z$
determines the degree of this additional non-exponentiality. Only
for $D_z\!\parallel\!\vec{\mu}$ the same stretching exponent as
for isotropic reorientation is obtained.
%
%
%

%
An entirely different approach to the calculation of RCFs via
rotational random walk simulations has proved to be successful in
the context of supercooled liquids\cite{gerald98}. Here, molecular
reorientation has been modelled by rotational jump processes.
Starting from an arbitrary orientation, after a certain waiting
time drawn from Poisson statistics the next orientation is chosen
at random with the only restriction that starting and end position
differ by an angle $\gamma$. In the limit of $\gamma \rightarrow
0$ isotropic rotational diffusion is obtained with results
identical to those obtained with the analytic approach.
\begin{figure}
  \includegraphics[width=6cm,height=7cm]{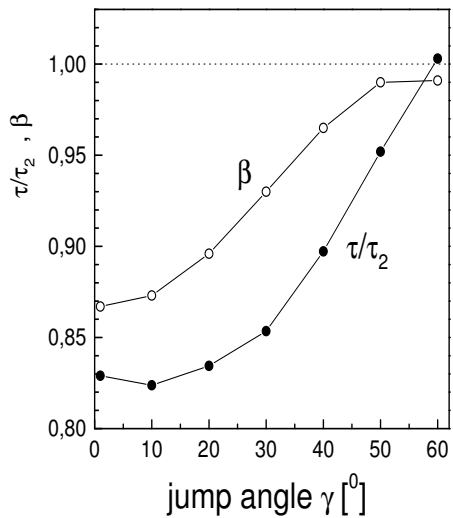}\\
  \caption{Influence of the reorientation mechanism on the
  stretching ($\circ$) and the time scale ($\bullet$) of the correlation
  function, calculated for rotational jump processes with varying
  jump angles $\gamma$. For comparison the correlation times were
  normalized by the pure $l=2$ correlation time $\tau_2$.
  }
\label{figrandomjump}
\end{figure}
In Fig.~\ref{figrandomjump} the influence of the jump angle
$\gamma$ on the correlation function $C_d(t)$ defined by
eqns.(\ref{def_d}) and (\ref{Cx.t}) is plotted. With increasing
$\gamma$ the correlation decays become more exponential, which
essentially originates from an convergence of the different $l$
contributions to the RCF.
%
%
%

%
So far, we have assumed that the principal axes system (P)
relevant for experimental observables has a fixed orientation
$\Om_{PD}$ relative to the 'diffusion tensor' system (D), and only
the orientation of the latter with respect to the laboratory
system (L) is time dependent: $\Om(t)=\Om_{DL}(t)$. In principle,
one can treat the case of internal rotations by assuming a
composite Markov process~\cite{vkamp81}
$\{\Om_{PD}(t),\Om_{DL}(t)\}$ and using an appropriate master
equation for the composite process. If the internal motion is
independent of the tumbling motion of the whole molecule one can
factorize the corresponding probability functions and averages.
This case is of particular relevance for the conformational
motions of polymers or proteins in solution and has been treated
extensively in the literature to which we refer\cite{WS78}.

{In the present analysis we have neglected the influence of the
numerical aperture (NA) on the optical path. Using very high NA
objectives (NA $>1$) out-of-plane contributions of the transition
(emission) dipole to the fluorescence signal have to be
considered\cite{ha1999,vdbout1}. However, it was pointed out that
even for NA$=1.25$ the dichroism signal is only slightly
influenced by this effect\cite{vdbout1}. Additionally, if one were
to measure rotational dynamics over an extended temperature range,
it is experimentally more convenient to use low NA ($<1$)
objectives. Here still an influence of the NA is expected, which,
however, becomes smaller the smaller the NA is. Nevertheless, the
accurate incorporation of the NA into our theoretical description
has still to be worked out. We also would like to mention that
methods have been developed which utilize the longitudinal field
component of high NA objectives to extract the full three
dimensional orientation of transition dipoles\cite{3D1,3D2}.
Although quite powerful, such measurements require a very good S/N
ratio for subsequent data analysis and are limited with regard to
time resolution. }

%

%
While in the past only bulk experimental techniques were available
for studying molecular dynamics, recent experimental progress has
enabled to study dynamical processes at the single molecule level.
In particular, fluorescence microscopy has allowed to track
molecular orientation in time. Here the merits of single molecule
microscopy emerge in probing rotational dynamics in heterogeneous
environments in a direct way.
We found that even for the model of isotropic rotational diffusion
the obtained correlation function decays non-exponentially.
Motional anisotropy increases the deviation from exponentiality.
If the RCFs calculated for the simple model of anisotropic
rotational diffusion are parametrized by a stretched exponential,
we find stretching parameters $\beta \geq 0.85$ for physically
reasonable values of the anisotropy $\delta$ between $0.3$ and
$3$. From this we conclude that only $\beta$ values smaller than
roughly $0.85$ can be taken as an indication for intrinsically
non-exponential dynamics. Our findings suggest a careful
interpretation of single molecule rotational correlation functions
as obtained by polarization resolved microscopy.

\end{document}